\documentstyle[aps,twocolumn,prl,psfig]{revtex}
\newcommand{\lsim}{\raisebox{1.5pt}{\small $<$}\hspace*{-6.7pt}\raisebox{-3pt}{\small $\sim$} }

\begin{document}
\draft
\title{Electric field induced suppression of  universal 
conductance fluctuations and dephasing in disordered systems}
\author{Arindam Ghosh}
\address{Department of Physics, Indian Institute of Science, 
Bangalore 560 012, India}
\author{A. K. Raychaudhuri}
\address{National Physical Laboratory, K.S. Krishnan Road, New Delhi 110 012, 
India}

\date{\today}

\twocolumn[\hsize\textwidth\columnwidth\hsize\csname 
@twocolumnfalse\endcsname

\maketitle
\begin{abstract}
We report a novel phenomenon that the universal conductance fluctuations 
(UCF) can be suppressed by a small electric field $E$. The experiment 
has been carried out on single crystals of Si doped heavily with P and B beyond
the critical composition of insulator to metal transition. The
phenomenon is identified as a consequence of electric field induced 
dephasing of the electron wavefunction. Over the range of measurements, 
the observed dephasing rate ($\tau_\phi^{-1}$) varied as $\tau_\phi^{-1}
= aT + bE^q$ with $q \approx 1.3$ and for $E \gg E^*$, a cross-over field,
$\tau_\phi^{-1} \sim E^q$, independent of $T$. This experiment also
establishes that the UCF can be utilized as a sensitive electron 
``interferometer'' to measure dephasing rate. 
\end{abstract}

]

Strong impurity scattering at low temperatures gives rise to a number of 
interesting phenomena due to quantum interference of the multiply back 
scattered electrons~\cite{tvr}. At $T =$ 0,
the electrical conductance $G$ of a disordered metallic system becomes 
a very sensitive function of the defect configuration and change in 
the position of even a single scatterer over a sufficient length scale 
($\sim k_F^{-1}, k_F$ is the Fermi wave vector) can produce a variation 
in the conductance $\delta G_1 \approx e^2/h$. This random, but 
reproducible variation in conductance with defect configuration, magnetic 
field or chemical potential is called the universal conductance 
fluctuations (UCF)~\cite{fls} and it arises due to interference of 
phase coherent electrons over large length scales.
At finite $T$, the length scale over 
which the interference is relevant is the phase coherence length $L_\phi$, 
which is related to the dephasing rate $\tau_\phi^{-1}$ through the 
relation $\tau_\phi^{-1} = D/L_\phi^2$, $D$ being the electron 
diffusivity. In the past decade and half experimental studies 
have established the occurrence of UCF in a number of 1D and 2D  disordered 
electronic systems~\cite{fls}. 
At low temperatures, UCF can be identified through
magneto-fingerprinting where the reproducible conductance variations 
are studied as a function of   
applied magnetic field $B$ in samples with one or more lateral dimensions             
\lsim $L_\phi$. In larger samples with dimensions $L \gg L_\phi$, 
UCF is observable as random time dependent conductance 
fluctuations with  approximately $1/f$ power spectra~\cite{ads2}. 

In  a recent experiment on heavily doped  
single crystals of Si (dopant P and B), it has been
shown by us that UCF can occur even in bulk 3D systems~\cite{agp1}. 
These single crystals with electronic concentration $n \approx 
(2-2.5)\times n_c$ where $n_c$ is the critical concentration for the 
insulator-metal transition, 
are disordered electronic system with $k_Fl \approx$ 2-5, $l$ 
being the elastic mean free path. They show such effects as weak 
localization and electron-electron interaction. The occurrence of the UCF in 
3D bulk crystals of Si has also been seen upto the range of Anderson 
transition ($n \approx n_c$)~\cite{agp2}.

In this paper we report a new and novel phenomenon where we were able to 
suppress the magnitude of UCF 
in these heavily doped Si single crystals by application 
of a small electric field. We explain this novel effect as arising from 
dephasing of the electrons by the applied electric field which in turn 
suppresses the UCF.

The phenomenon of UCF essentially rests on one crucial aspect, namely the 
phase coherence of the electron over a finite length scale $L_\phi \gg l$. 
Any interaction with the environment 
that introduces the phase decoherence, will also 
suppress the UCF. A very well known example is the suppression of the UCF 
by a factor of 2 in a magnetic field which breaks the time reversal        
symmetry~\cite{ads1}. 
The extreme sensitivity of the UCF to the phase coherence of the  
electron thus makes UCF a sensitive ``electron interferometer'' which can be 
used to measure dephasing of the electron. 
Our experiment is based on this basic concept.

We have carried out all our experiments in  single crystals of silicon
($\langle111\rangle$- Czochralski grown)  made 
metallic by doping  with P and B. (These are the same  samples 
on which we did the previous experimental work~\cite{agp1}.) The samples 
had dimensions of 0.5 mm $\times$ 0.10-0.15 mm and 
a thickness of $\approx 30 \mu$m. Sample volume for noise
detection ($\Omega$) $\approx 1.5-2.0\times10^{-12}$ m$^{3}$.
Noise, electrical conductivity and magnetoresistance (MR) were
all measured in the same sample to avoid any ambiguity. For noise
measurement we used a five probe ac technique~\cite{jsco}
aided by digital signal processing methods to measure extremely low 
magnitude of noise power (\lsim 10$^{-20}$ V$^2$/Hz).
The temperature stability was $|\Delta T/T| < 0.01\%$.
The Hall coefficient was found to
be essentially temperature independent down to 2 K with the
variation in the whole range being \lsim 20\%.

Experiments were done 
on a number of samples containing various concentration of P and B. They
yield qualitatively similar 
results. For simplicity and conciseness we report our findings on one of the 
samples. The sample (Si:P,B) contained P concentration of $1\times10^{25}$ 
m$^{-3}$. Disorder was introduced in the form of boron compensation 
(compensation factor $K \approx 0.4$). 
The net carrier concentration as obtained from Hall measurement is
$n \approx 2\times n_c$.  From the resistivity we find that  
$k_Fl \approx 3$ at low temperatures and hence the weak 
localization (WL) theories are applicable. At $T$ \lsim 10 K, the 
conductivity $\sigma(T)$ has a limiting correction $\Delta\sigma(T) 
\propto T^{m/2}$ arising from weak localization ($m \approx 1$). 
Magnetoresistance (MR) measurements showed additive contributions from
WL correction and electron-electron interaction. The phase breaking length 
$L_\phi$ was obtained from the WL contribution to MR. The details of 
the conductivity and MR analysis are given elsewhere~\cite{tvr}.

Noise measurements were done as a function of temperature and 
measuring electric field with  $E <$ 300 V/m. In figure~1 we show 
the measured relative fluctuation magnitude
$\langle\delta G^2\rangle/G^2$ as a function of $E$ for $T =$ 2 K. 
In the inset(a) of figure~1 we show the temperature dependence of the 
fluctuation measured with a very low field ($E \approx$ 7 V/m).
For $T$ \lsim 100 K the dominant contribution to the noise originates from 
the UCF mechanism as has been explained elsewhere~\cite{agp1}. 
Briefly, the rise of the fluctuation at low $T$ and the suppression 
of the noise by factor of 2 in a magnetic field (see inset (b) 
of figure~1) are 
the two hall marks for UCF. From figure~1 we see the important result 
that the noise is severely suppressed by even a moderate electric field.    
At $E \approx$ 250 V/m the noise is only about one fifth 
of that measured at 
$E \approx$ 10 V/m. It can be compared with the suppression of 
the noise by a magnetic field. In this sample, a magnetic field of about 
10$^{-2}$ T can suppress the noise by a factor of 2. 
A further suppression by another factor of 2 (due to removal of spin 
degeneracy) takes place at 
$H \geq$ 1.5 T. Thus the maximum suppression that one would get from the 
application of $H$ is a factor of 4. In contrast, in an electric 
field we have already achieved a suppression of factor of 5 and there is no 
approach to saturation detectable at the maximum electric field applied.

In  figure~2 we show the suppression of noise as a function of 
the electric field measured at three temperatures. It can be seen that the 
noise is suppressed by the electric field at all $T$. 
There is, however, a very 
interesting observation. Beyond a certain measuring field (marked  
$E^*(T)$), the measured  noise does not depend on $T$. 
Instead it is a function of the applied field $E$ alone. The 
cross over field $E^*(T)$ decreases as $T$ is decreased. 
We emphasize that what 
we are presenting here is the data in one sample. For various P and B 
concentration the data are qualitatively the same as long as the sample 
remains on the metallic side of the metal-insulator transition. In the 
following part of the paper we would like to provide an explanation of this 
experimental observation and would like to discuss it in the general 
perspective of the issue of decoherence in disordered system.

As stated earlier,
at $T = 0$, the conductance is an extremely sensitive function 
of the defect configuration. At finite but low
temperatures, as long as the phase coherence length $L_\phi \gg l$, 
this sensitivity is retained within a single phase coherent volume 
of $L_\phi^3$~\cite{ads2}. We had shown before from the experimentally
observed magnitude of conductance fluctuations in these systems that, 
$\langle(\delta G_\phi)^2\rangle$, the fluctuations in a  single
phase  coherent volume $L_\phi^3$, is actually saturated and has 
a value $\langle(\delta G_\phi)^2\rangle^{1/2} \approx 
1.5\times(e^2/h$)~\cite{agp1}.      
This is a very important observation and we will use it in our discussion 
below.

For a sample with  volume 
$\Omega \gg L_\phi^3$,
noise from different coherent regions of volume $L_\phi^3$ 
are superposed classically and the net relative conductance 
noise can be expressed as, 

\begin{equation}
\label{dgg}
\frac{\langle(\delta G)^2\rangle}{G^2} = \frac{L_\phi^3}{\Omega}\,
                                \frac{\langle(\delta G_\phi)^2\rangle}{G_\phi^2}
\end{equation}

\noindent where $G_\phi (= \sigma L_\phi)$ is the conductance 
of a single phase coherent box 
and $\sigma$ is the conductivity of the material. When the number of 
mobile/active scatterers in $L_\phi^3$ is sufficiently large, the mean     
square variance of conductance  saturates 
to $\langle(\delta G_\phi)^2\rangle^{1/2} \sim e^2/h$. 
As discussed before in this particular case the noise is indeed saturated 
and $\langle(\delta G_\phi)^2\rangle^{1/2} \approx 1.5\times(e^2/h)$.
In this case eqn.~\ref{dgg} can be simplified 
to

\begin{equation}
\label{dggt}
\frac{\langle(\delta G)^2\rangle}{G^2} \approx  
                   \frac{2.3L_\phi(T)}{\sigma^2\Omega}\,(e^2/h)^2 =
                   \frac{2.3\sqrt{D}(e^2/h)^2}{\sigma^2\Omega}\sqrt{\tau_\phi}
\end{equation}

\noindent Eqn.~\ref{dggt} clearly shows that the temperature dependence of 
saturated UCF noise is dominated by the temperature dependence of the phase 
coherence length $L_\phi = \sqrt{D\tau_\phi}$, or that of the dephasing 
rate $\tau_\phi^{-1}$. In fact one can utilize this information to obtain 
the value of the dephasing rate $\tau_\phi^{-1}$ in disordered systems at
low temperature~\cite{hc1}. 
Another way one can evaluate the dephasing rate is from 
the MR measurements. Since we have done both the measurements we can 
independently determine the dephasing rate $\tau_\phi^{-1}$. This has been 
shown in figure~3a. The measurements were done 
with an excitation electric field of 5 V/m. Within the experimental accuracy we 
found, $\tau_\phi^{-1} \propto T$. 
One can see that the dephasing rates determined from 
both the measurements agree quite well. This particular check of internal 
consistency establishes clearly that it is $\tau_\phi$ that predominantly  
determines the temperature dependence of the observed noise.

We  argue that the suppression of the noise in the electric field 
arises due to increase of the dephasing rate (hence reduction of 
$\tau_\phi$) in an applied electric field. 
In systems with strong electron-electron interaction it has been 
shown that such a low frequency electric field may in fact cause 
dephasing in the particle-hole channel~\cite{mlb}. 
This effect occurs when two interacting electrons
moving in the same closed Feynman path releases an excitation 
of energy $\epsilon$ at some instant $t^\prime$ and traverse rest 
of the path with unequal momentum under 
an ambient time dependent vector potential. Quantitatively, the phase 
difference acquired in such a process 
$\Delta\phi/2\pi = e \Delta{\bf x}.{\bf E}\eta$, where 
$\Delta{\bf x}$ is the displacement between the point of interaction 
and that of observation and $\eta$ is time taken to traverse the full path. 
The interference is lost when $\Delta\phi/2\pi \approx 1$ within a 
thermal path length $L_T =
\sqrt{\hbar D/k_BT}$. Since only those paths with $\eta \leq \hbar/k_BT$
contribute to the phase relaxation~\cite{alt1}, $\Delta\phi/2\pi \approx 1$
when
$e E L_T \approx k_BT$. This condition defines an energy scale $\Sigma(E)$
defined as 

\begin{equation}      
\label{es}
\Sigma(E) = (\hbar e^2DE^2)^{1/3}
\end{equation} 

\noindent We will discuss the implication of the energy scale $\Sigma(E)$   
later on.

If indeed the electric field introduces         
dephasing (i.e, increases $\tau_\phi^{-1}$) and hence suppresses the noise,  
then the electric field dependence of the measured noise  can be converted 
into a dependence of $\tau_\phi$ on $E$ using eqn.~\ref{dggt}.
In the inset of figure 2 we show the value of 
$\tau_\phi^{-1}$, as a function of the electric field $E$.
At low electric fields, $\tau_\phi^{-1}$ is independent of $E$, as 
expected of a linear system. As the field is increased, $\tau_\phi^{-1}$ 
increases implying an increase in the total dephasing rate. By scanning 
over 2 orders of 
magnitude of electric field, the dephasing rate increases by similar order.

We observe that at a large enough $E$ ($E \gg E^{*}$)
the value of $\tau_\phi$ becomes independent of $T$ and 
depends essentially on $E$. In this regime
the dephasing rate $\tau_\phi^{-1} \propto E^q$, where $q \approx 
1.3\pm0.05$.
For small $E$ ($E \ll E^*$), the dephasing rate $\tau_\phi^{-1}
\propto T^p$, where $p \approx 1.0\pm0.05$. We can 
interpolate in the intermediate field region  using the relation:

\begin{eqnarray}
\label{ttphi}
\tau_\phi^{-1} & = & aT^p + bE^q 
\end{eqnarray} 
\noindent where $a$ and $b$ are constants independent of $E$ and $T$.
The fit of the dephasing rate data to eqn~\ref{ttphi} are shown in 
the inset of figure~2. This particular way of expressing the dephasing rate 
assumes that we have two independent dephasing channels. The temperature 
dependent part arises from the usual inelastic scattering 
(e.g, the electron-electron interaction, electron-phonon interaction or 
TLS-electron interaction etc.)~\cite{alt1,2ck,TLS}. 
The field dependent part of the 
dephasing is expected to be directly related to $\Sigma(E)$, the energy 
scale that characterizes the extra phase the electron gains from the 
field $E$. We believe this dephasing is a many-body effect 
arising from the
electron-electron interaction. For such a process, the quasiparticle       
scattering rate ($\tau_{ee}^{-1}$) depends on the energy transfer in 
the process ($\epsilon$) 
and one can formally write $\tau_{ee}^{-1} \propto \epsilon^\zeta$. We 
suggest that in this particular case of field induced dephasing the energy 
transfer will be determined by $\Sigma(E)$ (see eqn~\ref{dggt}) so 
that $\epsilon \sim \Sigma(E)$. We will 
then have $\tau_{ee}^{-1} \propto E^{2\zeta/3}$. 
Experimentally, $q \approx 1.30\pm0.05$ so
that $\zeta \approx 1.95\pm0.07$. From the Fermi liquid theory for a 
clean system with long 
mean free path $\zeta = 2$ and for a dirty systems with short mean free 
path $\zeta \approx 1.5$~\cite{alt1}. 
The value of $\zeta$ estimated from the experiment is thus quite 
close to what is expected from this simple theoretical approach. 

We can define a temperature dependent cross-over field $E^*(T)$ from our 
experiment using eqn~\ref{ttphi} so that when $E = E^{*}$ 
the dephasing rate 
obtained from both channels are equal. We obtain $E^{*} \propto T^{p/q}
\approx T^{1/q}$. In figure~3b we show the variation of observed $E^*$ as 
a function of $T$. The solid curve is the line $T^{1/q}$ where 
$q \approx$ 1.3 as obtained from the experiment. The agreement is very 
good.

We have the following strong  reasons to believe that this effect is not 
due to heating of the sample in the usual sense: 
(1) even at the highest bias the power dissipation is \lsim 20 
$\mu$W, (2) the value of $\tau_\phi$ obtained at the highest bias 
corresponds to that at
$T \approx 200$ K if the complete dephasing was due to electron heating, 
which is rather unlikely when the sample is held 
at 2 K and
(3) if there would have been electron heating, the conductivity $\sigma$
would have   
been strongly affected and would have become a strong function of the 
field. The observed field dependence of the conductivity then can be used
as a ``thermometer'' for the electron temperature. We find a small 
dependence 
of $\sigma$ on the electric field. At the highest field and the lowest 
$T$, $\delta\sigma/\sigma$ \lsim 0.1\%. Taking $\sigma$ as the temperature 
scale and assuming that the entire field dependence of $\sigma$
arises from heating we obtain an upper limit of the electron  temperature 
rise of $\sim$ 0.05 K. We hence conclude that the dephasing seen with 
applied electric field is not a heating effect.
                                    
Our experiment thus revealed for the first time that at low 
temperature dephasing can be induced by an $E$ field. In none 
of the earlier studies, made in low dimensional systems
like wires, films etc.~\cite{est}, 
such a field induced dephasing was reported. 
We make use of the UCF as a direct probe to find the 
dephasing, unlike the previous  
experiments where $\tau_\phi$ was extracted as a fit parameter 
from MR experiments. Given the sensitivity of the UCF 
to the phase of electron wave functions, it may actually be a 
better tool for detecting dephasing.

In recent years there is an interesting debate on the issue of electron
dephasing at low temperatures~\cite{mjw}. In this context our experiment 
can be seen as a useful and new contribution. 
Using UCF as a sensitive probe of dephasing, 
we establish that the measuring field can 
induce dephasing (without electron heating) in disordered systems with
interacting electrons. At lower temperatures, such phase relaxation is 
brought about by relatively smaller fields and beyond a characteristic
field scale $E^*(T)$, the dephasing rate becomes independent of 
temperature and is determined only by the electric field. 

We thank Prof. D.F. Holcomb of Cornell University for providing 
the samples used in this experiment.

{\bf\large Figure caption:}

{\bf figure 1:} Electric field dependence of total noise magnitude at 2 K.
The solid line is the fit as described in text. Inset (a) shows the 
temperature dependence of the noise magnitude measured with a field of 
$\approx$ 7 V/m. Inset (b) shows the magnetic field dependence of noise.
$\nu(H) = \langle(\delta G)^2_H\rangle/\langle(\delta G)^2_{H=0}\rangle$. 
Two separate reductions of $\nu(H)$ by factors of 1/2 clearly 
establishes the signature of UCF.

{\bf figure 2:} Electric field dependence of noise at three different 
temperatures. Arrows denote the cross-over field $E^*$. The inset 
shows the similar dependence of the dephasing rate $\tau_\phi^{-1}$ 
obtained using eqn.~\ref{dggt}. The dotted lines are fits as described 
in text.

{\bf figure 3:} (a) Temperature dependence of $\tau_\phi$  from noise 
(eqn.~\ref{dggt}) and MR measurements. The measuring field $E \ll E^*$.
Both the lines have slope $\approx 1.0\pm0.05$.
(b) Temperature dependence of the cross-over field $E^*$. The slope of 
the line is $\approx$ 0.75 which is fairly close to the expected value
of $p/q \approx 0.77$.

\begin{figure}[htb]
\centering \leavevmode
\psfig{file=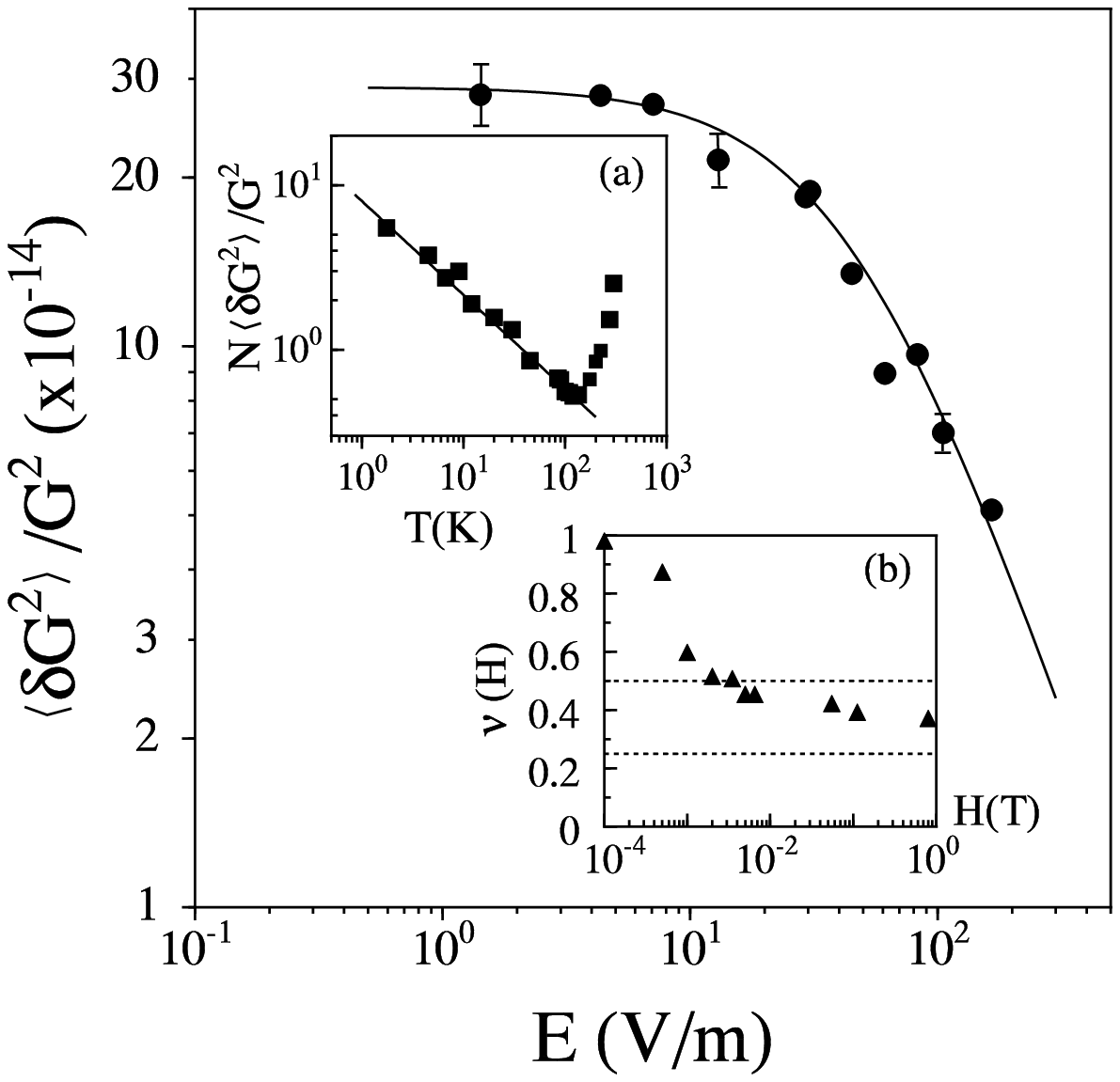,width=0.725\hsize}
\end{figure}

\begin{figure}[htb]
\centering \leavevmode
\psfig{file=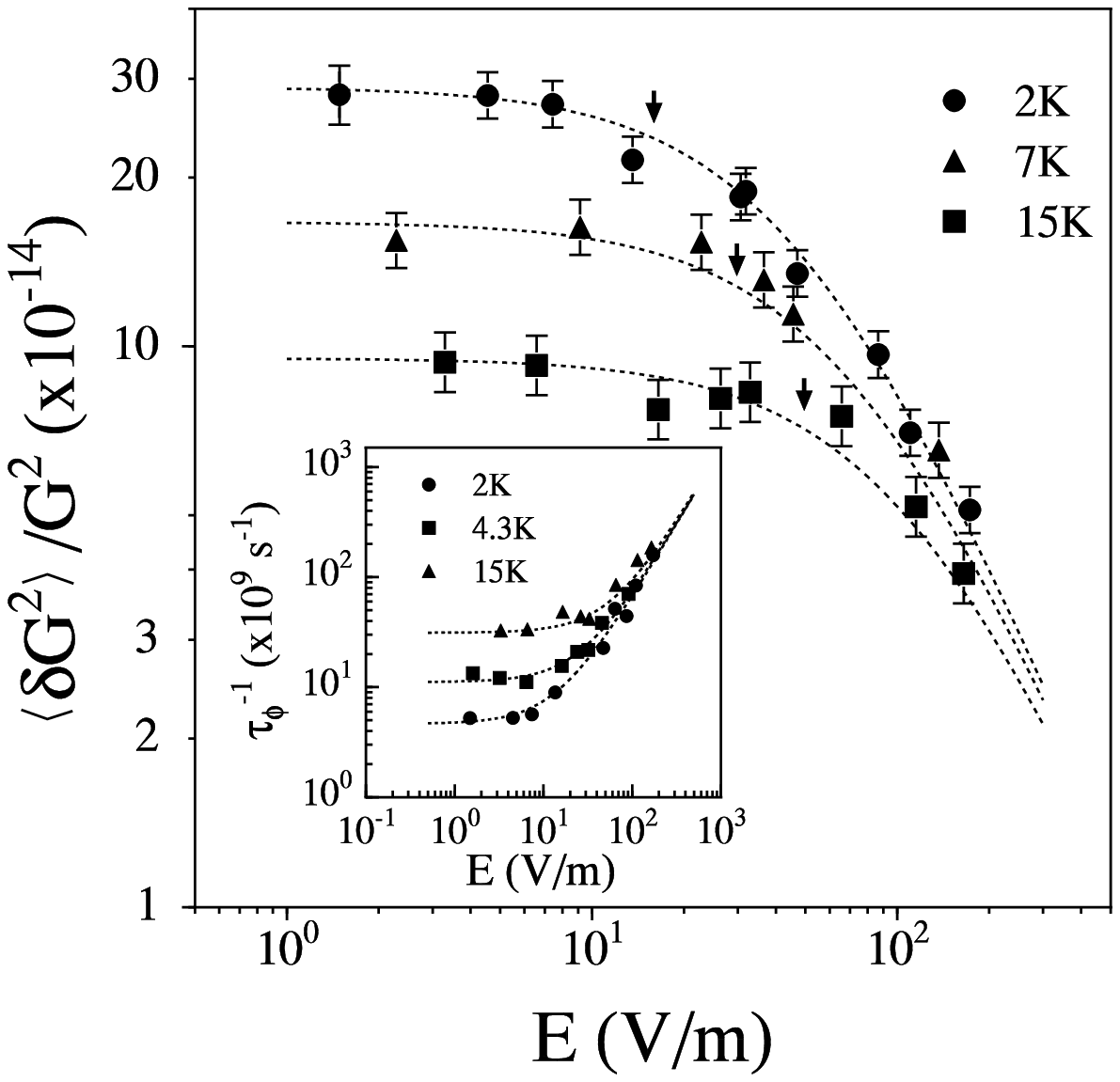,width=0.725\hsize}
\end{figure}

\begin{figure}[htb]
\centering \leavevmode
\psfig{file=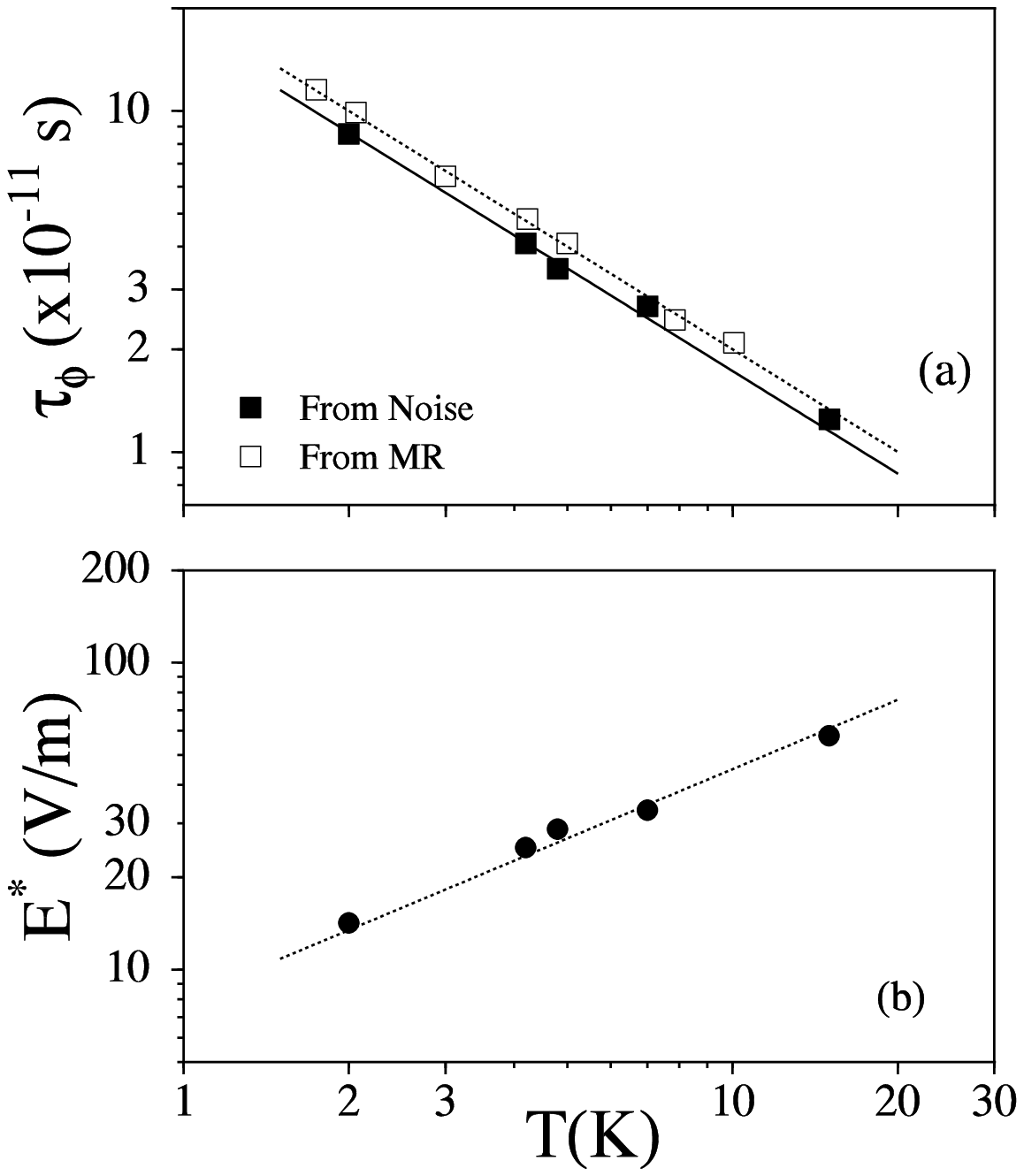,width=0.725\hsize}
\end{figure}


\begin{references}
\bibitem{tvr} P.A. Lee and T.V. Ramakrishnan, Rev. Mod. Phys. {\bf 57},
287 (1985)
\bibitem{fls} P.A. Lee, A.D. Stone and H. Fukuyama, Phys. Rev. B {\bf 35},
1039 (1987); N. Giordano, in {\it Mesoscopic Phenomena in Solids}, {\it ed.}
B.L. Altshuler, P.A. Lee and R.A. Webb, Elsevier Sc. Publ. B.V. (1991)
\bibitem{agp1} Arindam Ghosh and A.K. Raychaudhuri, cond-mat/9907158
\bibitem{agp2} Arindam Ghosh and A.K. Raychaudhuri, J. Phys.: Condens. Matter 
{\bf 11}, L457 (1999)
\bibitem{ads1} A.D. Stone, Phys. Rev. B {\bf 39}, 10736 (1989)
\bibitem{jsco} J.H. Scofield, Rev. Sci. Instrum.  {\bf 58}, 985 (1987).
\bibitem{ads2} S. Feng, P.A. Lee and A.D. Stone, Phys. Rev. Lett. {\bf 56}, 
1960 (1986)
\bibitem{hc1} The usual method of determining $L_\phi$ from UCF is from the
magnetic field scale of the cross-over function. However, this method requires
application of magnetic field to break the time reversal symmetry and is 
analogous to WL.
\bibitem{mlb} M. Leadbeater, R. Raimondi, P. Schwab and C. Castellani,
cond-mat/9905240
\bibitem{alt1} B.L. Altshuler and A.G. Aronov, in {\it Electron-Electron 
Interaction in Disordered Systems}, edited by A.L. Efros and M. Pollak 
(North Holland, Amsterdam, 1985)
\bibitem{2ck} A. Zawadowski, Jan von Delft and D.C. Ralph, cond-mat/9902176
\bibitem{TLS} The temperature dependent dephasing with $p\approx 1$ most 
likely arises from electron TLS interaction~\cite{2ck}. The other 
conventional processes give rise to $p\geq 2$.
\bibitem{est} N.O. Birge, B. Golding and W.H. Haemmerle, Phys. Rev. B
{\bf 42}, 2735 (1990); J.S. Moon, N.O. Birge and B. Golding, {\it ibid}
{\bf 56}, 15124 (1997); G.A. Garfunkel, G.B. Alers, M.B. Weissman, 
J.M. Mochel and D.J. VanHarlingen, Phys. Rev. Lett. {\bf 60}, 2773 (1988)
\bibitem{mjw} P. Mohanty, E.M.Q. Jariwala and R.A. Webb, Phys. Rev. Lett. 
{\bf 78}, 3366 (1997); I.L. Aleiner, B.L. Altshuler and M.E. Gerhenson, 
cond-mat/9808053; Y. Imry, H. Fukuyama and P. Schwab, Europhys. Lett. 
{\bf 47}, 608 (1999)
\end{references}
\end{document}